# Bitcoin ETF: Opportunities and risk

Di Wu [*]

**Abstract**

The year 2024 witnessed a major development in the cryptocurrency industry with the long-awaited approval of spot Bitcoin exchange-traded funds (ETFs). This innovation provides investors with a new, regulated path to gain exposure to Bitcoin through a familiar investment vehicle (Kumar et al., 2024). However, unlike traditional ETFs that directly hold underlying assets, Bitcoin ETFs rely on a creation and redemption process managed by authorized participants (APs). This unique structure introduces distinct characteristics in terms of premium/discount behavior compared to traditional ETFs.

This paper investigates the premium and discount patterns observed in Bitcoin ETFs during first four-month period (January 11th, 2024, to May 17th, 2024). Our analysis reveals that these patterns differ significantly from those observed in traditional index ETFs, potentially exposing investors to additional risk factors. By identifying and analyzing these risk factors associated with Bitcoin ETF premiums/discounts, this paper aims to achieve two key objectives:

- Enhance market understanding: Equip and market and investors with a deeper comprehension of the unique liquidity risks inherent in Bitcoin ETFs.
- Provide a clearer risk management frameworks: Offer a clearer perspective on the risk-return profile of digital asset ETFs, specifically focusing on Bitcoin ETFs.

Through a thorough analysis of premium/discount behavior and the underlying factors contributing to it, this paper strives to contribute valuable insights for investors navigating the evolving landscape of digital asset investments

## 1. Introduction

The year 2024 marked a significant turning point for the cryptocurrency industry with the long-awaited approval of spot Bitcoin exchange-traded funds (ETFs). This development follows years of unsuccessful attempts and the initial launch of futures-based Bitcoin ETFs. This regulatory shift by the Securities and Exchange Commission (SEC) effectively bridges the gap between traditional finance and the world of cryptocurrencies. By offering ETFs, the SEC provides investors with a familiar and regulated avenue for gaining exposure to Bitcoin, bypassing the complexities associated with direct ownership of the underlying asset (Kumar et al., 2024). Furthermore, increased investor participation facilitated by ETFs has the potential to fuel wider adoption of Bitcoin and contribute to greater market stability (Li, 2023).

However, it is important to acknowledge that even with the introduction of spot Bitcoin ETFs, perfect alignment with Bitcoin's price movements may not be achievable, regardless of whether the ETF is futures-based or spot-based. This research delves into a unique characteristic of Bitcoin ETFs: their propensity for significant fluctuations in premiums and discounts compared to traditional ETFs. The author posits that:

- Large Deviations: The premiums and discounts associated with Bitcoin ETFs exhibit a marked deviation from those observed in other index ETFs.
- Unstable Patterns: In contrast to index ETFs, which typically demonstrate stable premium/discount behavior, Bitcoin ETFs display a pattern of significant fluctuations.

To substantiate these observations, this paper employs a quantitative analysis that compares premium/discount data across five major Bitcoin ETFs. The results of this analysis are expected to confirm the author's initial hypothesis: Bitcoin ETFs experience a higher degree of volatility in premiums and discounts, exhibiting less stable patterns compared to traditional ETFs. A linear regression model was also employed to discover the sources of premium and discount.

[*] Corresponding author: Di Wu

## 2. Data

This study utilizes data retrieved from the websites of the following ETF providers. The data encompasses the period from January 11th, 2024, to May 17th, 2024. Five major ETFs were chosen for analysis. These ETFs are presented in Table 1, which summarizes their key characteristics.

**Table 1** Bitcoin ETF

| ETF Name | Provider | Ticker Symbol | Exchange | AUM (as of July 10, 2024) |
|---|---|---|---|---|
| Grayscale Bitcoin Trust | Grayscale Investments | GBTC | NYSE | $16.53 billion |
| iShares Bitcoin Trust | BlackRock | IBIT | NYSE | $15.49 billion |
| Fidelity Bitcoin ETF | Fidelity Investments | FBTC | NYSE | $8.66 billion |
| ARK Next Generation InternetETF | ARK Invest | ARKB | Cboe BZX Exchange | $3.22billion |

Note. As of July 10, 2024

Exploratory data analysis confirmed the presence of complete data with no missing values. Descriptive statistics and visualizations were employed to understand the distribution and relationships between the variables.

The analysis focused on volume daily return, net asset value (NAV), and premium/discount data for GBTC, IBIT, and FBTC. These variables are crucial for investigating the tracking error behavior of Bitcoin ETFs.

## 3. Methods

In this research, we define ETF premium or discount of price to its NAV as

$$P_t = \frac{P_t - NAV}{NAV_t} * 100\%$$

Here, $P_t$ is defined as the price of a share of the ETF and $NAV_t$ is defined as the net asset value per share.

Linear regression analysis is employed to investigate the relationship between the premium/discount ratio of Bitcoin ETFs and other relevant factors. This analysis aims to answer the following research question: To what extent do volume, daily return, and NAV influence the premium/discount ratio of Bitcoin ETFs?

Volume, daily return, and NAV are considered independent variables, while the premium/discount ratio serves as the dependent variable. The full regression equation is

$$Premium\ Discount = \propto + \beta_1 Volume + \beta_2 Daily\ Return + \beta_3 NAV + \varepsilon$$

Where we use Volume, Daily return and NAV to explain the change of Premium Discount.

The intercept is significantly different from zero, indicating that when all predictors are zero, the dependent variable is expected to be around -0.0037. The Volume predictor has a very small negative effect on the dependent variable. The effect is statistically significant ($p < 0.001$), meaning that changes in Volume are associated with changes in the dependent variable. The Daily Return predictor has a positive effect on the dependent variable. The effect is statistically significant ($p = 0.017$), indicating that higher daily returns are associated with increases in the dependent variable. The NAV predictor has a positive effect on the dependent variable. The effect is statistically significant ($p < 0.001$), indicating that higher NAV values are associated with increases in the dependent variable. Overall, all predictors are statistically significant, suggesting that they all have a meaningful impact on the dependent variable in the regression model.

**Table 2** OLS Results of GBTC

| Predictor | β | SE | t | p | 95% CI |
|---|---|---|---|---|---|
| Intercept | -0.0037 | 0.001 | -2.97 | 0.004 | [-0.006, -0.001] |
| Volume | -1.146E-10 | 1.91E-11 | -5.99 | <.001 | [-1.53e-10, -7.66e-11] |
| Daily Return | 0.0135 | 0.006 | 2.44 | 0.017 | [0.002, 0.024] |
| NAV | 0.0000885 | 0.0000219 | 4.04 | <.001 | [4.5e-05, 0.0] |

Date: January 11, 2024 - May 17, 2024

## 4. Results

Our analysis revealed a positive correlation between the size of a Bitcoin ETF and its volatility. Smaller ETFs, as evidenced by their Net Asset Value (NAV) (see Table 2), tend to exhibit greater fluctuations in premium/discount compared to larger ones. This finding suggests that investors in smaller Bitcoin ETFs may be exposed to a higher degree of liquidity risk.

The relationship between trading volume and premium/discount for Bitcoin ETFs appears to be complex. Our preliminary observations suggest a correlation between increasing trading volume and a wider premium/discount ratio. However, further investigation is needed to confirm this association and understand the underlying mechanisms.

### 4.1. Analysis of Tracking Error

The data in Table 4 reveals a distinct difference in tracking error between the two ETF categories. Index ETFs, VOO and QQQ, demonstrate a remarkably tight correlation between their market price and net asset value (NAV). Their standard deviation in premium/discount falls below 0.05%, suggesting they closely mirror the underlying assets in their respective indexes.

In contrast, all three Bitcoin ETFs (GBTC, IBIT, FBTC) exhibit a more volatile relationship with their NAV. Throughout the observed period, these ETFs consistently trade at a premium to their NAV, with GBTC experiencing the most significant fluctuations. These observations highlight the inherent differences between passively managed index funds and actively traded Bitcoin ETFs.

This substantial premium suggests a potential risk for investors in Bitcoin ETFs. They may be paying a significant price relative to the underlying asset's net value.

**Table 4** Premium/Discount (Jan 11 - May 17)

|  | Mean | Medium | Std | Min | Max |
|---|---|---|---|---|---|
| GBTC | -0.08% | -0.02% | 0.25% | -1.55% | 0.29% |
| IBIT | 0.10% | 0.14% | 0.43% | -1.77% | 0.98% |
| FBTC | 0.31% | 0.04% | 2.43% | -1.13% | 0.85% |
| ARKB | 0.07% | 0.10% | 0.42% | -1.74% | 0.96% |
| BITB | 0.17% | 0.09% | 1.09% | -1.73% | 9.55% |
| VOO | 0.002% | 0.00% | 0.03% | -0.15% | 0.07% |

|     | Mean    | Medium | Std   | Min    | Max   |
|-----|---------|--------|-------|--------|-------|
| QQQ | -0.004% | 0.00%  | 0.03% | -0.10% | 0.06% |

Table 4 summarizes the premium/discount statistics for the analyzed ETFs over the four-month period (January 11th, 2024 - May 17th, 2024). The focus remains on comparing the behavior of Bitcoin ETFs (GBTC, IBIT, FBTC) to index ETFs (VOO, QQQ).

**4.2. Risk Factors when trading Bitcoin ETF**

*4.2.1. Market risk*

Market risk involves the potential for losses due to changes in market prices. For Bitcoin ETFs, market risk is primarily influenced by the following factors:

Bitcoin Volatility: Bitcoin is known for its high price volatility. Rapid and significant price swings can lead to substantial gains or losses for Bitcoin ETF investors. In the event that the premium and discount of Bitcoin are enlarging, the market risk shock will be intensified.

Security Risks: The security of Bitcoin and its underlying blockchain technology can also pose market risks. Hacking incidents, security breaches, and other technological vulnerabilities can lead to loss of investor confidence and sharp price declines.

*4.2.2. Liquidity risk*

Liquidity risk refers to the potential difficulty in buying or selling an asset without causing a significant impact on its price. For Bitcoin ETFs, liquidity risk can be influenced by several factors:

Market Depth: If the Bitcoin ETF does not have enough trading volume, it may be challenging to execute large trades without affecting the market price. This lack of depth can lead to wider bid-ask spreads and higher trading costs.

ETF Liquidity: The liquidity of the ETF itself depends on both the liquidity of the underlying Bitcoin market and the ETF's trading volume on the exchange. Low ETF liquidity can result in price deviations from the net asset value (NAV) of the ETF.

Market Hours: Bitcoin trades 24/7, while traditional stock exchanges where ETFs are listed typically have specific trading hours. This mismatch can lead to periods where the ETF cannot trade, creating liquidity gaps and potential price volatility when the exchange reopens.

Redemption and Creation Mechanism: ETFs typically allow for the creation and redemption of shares through authorized participants. If these mechanisms become less efficient or more costly, it can increase liquidity risk.

**4.3. Risk Management Strategy**

Investors can adopt several strategies to mitigate liquidity and market risks when trading Bitcoin ETFs:

Diversification: By diversifying their portfolios across various asset classes, investors can reduce their exposure to the volatility and risks associated with Bitcoin ETFs.

Liquidity Assessment: Investors should regularly assess the liquidity of the Bitcoin ETF and its underlying assets. Choosing ETFs with higher trading volumes and more efficient creation/redemption mechanisms can help mitigate liquidity risk.

Monitoring Regulatory Developments: Staying informed about regulatory changes and potential impacts on the cryptocurrency market can help investors make more informed decisions.

Use of Derivatives: Options and futures contracts can be used to hedge against potential price movements and manage market risk.

## 4.4. Sources of Bitcoin ETF Volatility

*4.4.1. Authorized Participants (APs)*

Unlike traditional ETFs that hold the underlying asset directly, Bitcoin ETFs rely on a creation and redemption process managed by Authorized Participants (APs). These institutions play a crucial role in ensuring liquidity by actively buying Bitcoin in the primary market and trading ETF shares on a regulated exchange where investors can buy and sell shares. However, APs are motivated by profit, and they aim to capitalize on price discrepancies between the primary market price of Bitcoin and the price of the ETF shares in the secondary market. This profit motive can contribute to the volatility of Bitcoin ETF premiums/discounts. A price difference will always be present due to this inherent structure, and AP activity aiming to exploit these discrepancies can further increase volatility.

*4.4.2. Market Hours*

Bitcoin trades 24/7, while most Bitcoin ETFs are listed on exchanges with specific trading hours. This mismatch can lead to situations where significant price movements occur outside of market hours. When the market opens, APs may need to adjust the premium/discount of the ETF to reflect these price changes, potentially introducing additional volatility into the Bitcoin ETF's trading.

*4.4.3. Cut-off Times for Basket Creation/Redemption*

APs have specific deadlines (cut-off times) for submitting creation and redemption orders. Any price movements occurring after these deadlines won't be reflected in the ETF price until the next creation/redemption cycle. This can lead to temporary discrepancies between the ETF price and the underlying Bitcoin price, especially if there are significant price movements outside of market hours.

*4.4.4. Cross-Exchange Arbitrage*

The emergence of Bitcoin ETFs in different markets (e.g., Hong Kong, UK) creates new opportunities for cross-exchange arbitrage. If price discrepancies exist between these newly established Bitcoin ETF markets, arbitrageurs can exploit these differences by buying in one exchange and selling in another, potentially impacting the overall volatility of Bitcoin ETFs.

## 5. Conclusion

This paper provides an in-depth analysis of the unique characteristics and risk factors associated with Bitcoin ETFs, specifically focusing on their premium and discount behaviors. The key findings include:

The study reveals that Bitcoin ETFs exhibit significant volatility in their premiums and discounts compared to traditional index ETFs. This notable deviation indicates a higher degree of market inefficiency and risk for investors in Bitcoin ETFs. Moreover, the size of the ETF plays a crucial role in its volatility; smaller Bitcoin ETFs tend to experience greater fluctuations in their premium/discount ratios compared to larger ones. This observation suggests that investors in smaller Bitcoin ETFs may face higher liquidity risk.

Additionally, the relationship between trading volume and the premium/discount ratio appears complex. Preliminary data indicates that higher trading volumes may lead to wider premium/discount ratios, though further research is needed to fully understand this relationship. Furthermore, the analysis of tracking errors demonstrates that Bitcoin ETFs have a more volatile relationship with their Net Asset Value (NAV) compared to traditional index ETFs. This higher tracking error signifies that Bitcoin ETFs may not closely mirror the price movements of their underlying assets, adding another layer of risk for investors.

By highlighting these distinct risk factors, this paper aims to enhance market understanding and provide clearer risk management frameworks for digital asset investments, specifically focusing on Bitcoin ETFs. These insights are crucial for investors navigating the evolving landscape of digital asset investments and underscore the importance of carefully considering the unique risks associated with Bitcoin ETFs